\documentclass[12pt]{article}
\usepackage{a4,epsf,times,amssymb}

\newcommand{\Ss}[1]{{\sf\small #1}}

\begin{document}
\vspace{1.3in}
\begin{center}
{\Large {\bf Distributed Computation as Hierarchy}}\\
\vspace{0.2in}
{\large Michael Manthey}\\
{\small Computer Science Department, Aalborg University\\Fr. Bajersvej 7E; 9200 
Aalborg, Denmark\\manthey@cs.auc.dk}\\
\copyright \today.
\end{center}

{\small
\begin{itemize}
\item[] {\bf Abstract}.
This paper presents a new distributed computational model of distributed systems 
called the {\em phase web} that extends Vaughn Pratt's orthocurrence relation 
from 1986. The model uses mutual-exclusion to express sequence, and a new kind 
of hierarchy to replace event sequences, posets, and pomsets. The model 
explicitly connects computation to a discrete Clifford algebra that is in turn 
extended into homology and co-homology, wherein the recursive nature of objects 
and boundaries becomes apparent and itself subject to hierarchical recursion. 
{\em Topsy}, a programming environment embodying the phase web, is currently 
being readied for release.

\item[] {\bf Keywords}: process, hierarchy, distributed, co-occurrence, 
co-exclusion, orthocurrence, Clifford algebra, homology, co-homology, twisted 
isomorphism, phase web paradigm, Phase Web, Topsy, reductionism, emergence, 
Kron.

\end{itemize}
}
 
\section*{Introduction}

The intent of this paper is to introduce a new model of distributed computation, 
and presumes that the reader is familiar with extant models (CCS, pomsets, and 
the like) so as to concentrate on what is new about the present approach. The 
model arose in the author's search for a way to describe and specify a 
particular class of distributed computations, namely self-organizing systems, 
although it is applicable to less demanding applications as well. Prominent 
characteristics of self-organizing systems are (1) they are not intended to 
`halt', (2) they are meaningless when separated from their environment, with 
which they constantly interact, (3) they must be self-reflective, and thus (4) 
they are, in a non-teleological sense, goal-driven. [Man97] enlarges on these 
themes.

In contrast to many, the approach presented here emphasizes {\em structure} so 
strongly that the algorithmic component that for most people is the sine qua 
non of computation is nearly invisible.  This emphasis is ultimately the reason 
why the approach offered here - called {\em the phase web paradigm} - differs 
from all others we are familiar with, and correspondingly, why its mathematics 
comes out so differently (algebraic topology, namely, rather than logic).

Of course one still writes programs, but in pure process-coordination terms; we 
use a local extension to Linda [Gel85] called TLinda [Note 0]. However, since a 
self-organizing system generally grows/learns, this programming is ultimately 
sculptural rather than specificational in character. This sculpturing is a 
reflection of the hierarchical aspect of the phase web.

As will become apparent, this work adopts algebraic topology as its overall 
mathematical framework. The first to do this were Herlihy and Shavit [HS93], who 
model decision tasks via simplicial complexes, and then use homology theory to 
capture state change. We use only the former, and, unlike these authors, in a 
way that is quite independent of the application. Nevertheless, we are very much 
in agreement with Herlihy and Rajsbaum when they write [HR95] ``We believe that 
these techniques and models, borrowed from classical algebraic topology, 
represent a promising new approach to the theory of distributed computing. 
Because these notions come from a mature branch of mainstream mathematics, the 
terminology (and to a lesser degree, the notation) is largely standardized, and 
the formalisms have been thoroughly debugged. Most importantly, however, this 
model makes it possible to exploit the extensive literature that has accumulated 
in the century since Poincar\`e and others invented modern algebraic topology''.

Also related is Pratt's work with Chu spaces and automata [Pratt95]. Automata 
are a traditional way to view computation, and state and prove formal 
properties, although their very sequentiality is problematic when dealing with 
distributed computations. It is therefore interesting that the functionality of 
a Chu automaton appears to have a counterpart in the $\mu,\mu^{-1}$ mappings of 
the ladder hierarchy we present here.

The next section introduces the basic computational insight, at the end of which 
we give an outline of the remainder of the paper.

\section{Pomsets and vector products}

In a remarkable passage, [Pratt86] introduces, in the context of discrete events 
and pomsets thereof, the seed concept as follows.

{\small
\begin{itemize}
\item[] The mental leap from strings to pomsets has much in common  with that 
from reals to complex numbers. In particular, one can encounter complex numbers 
without previously having seen a definition of the concept, via operations on 
real numbers, the canonical example being to apply square root to a negative 
number. A similar situation obtains with pomsets, as the Figure below 
illustrates in the application of ``pomset multiplication'' to two strings. In 
this figure, vertices denote pomset events, and the order on the events is given 
by the reflexive transitive closure of the relation implied by the arrows.
\end{itemize}

$$
\begin{array}{cccccccccccccccc}
0&&T&&(0,T)&\longrightarrow&(0,R)&\;&\;&\;&0&&T&&0&T\\
\\
\downarrow&\times&\downarrow&=&\downarrow&&\downarrow&&&& 
\downarrow&||&\downarrow& = &\downarrow&\downarrow\\
\\
1&&R&&(1,T)&\longrightarrow&(1,R)&&&&1&&R&&1&R\\
\\
&&&&(a)&&&&&&&&(b)
\end{array}
$$

\begin{center}
(a) Orthocurrence and (b) concurrence, of strings $01$ and $TR$.
\end{center}

\begin{itemize}
\item[] In the figure (a), we have ``multiplied'' two strings $01$ and $TR$ to 
yield not a string but a poset. The multiplication, which we call orthocurrence, 
may be characterized as a direct product of partial orders. This multiplication 
has a very natural and useful interpretation: if we interpret $TR$ as modelling 
the sequence Transmit-then-Receive, and $01$ as the message $0$ followed by the 
message $1$, then the product is immediately recognizable as the four events 
Transmit 0, Receive 0, Transmit 1, Receive 1. Furthermore the order is equally 
recognizable as the necessary temporal order on these four events! We shall call 
pomset multiplication {\em orthocurrence}.

\item[] The need for pomsets rather than posets becomes apparent when we 
substitute the string $00$ for $01$ in this example. We then have four events 
consisting of two occurrences of each of the actions Transmit 0 and Receive 0. 
This constitutes a multiset with four elements but with only two distinct 
elements.

\item[] Because of this possibility of repetitions of actions we draw a 
distinction between events and actions. Actions label events, that is, there is 
a labelling function from events to actions. An event is an instance or 
occurrence of its action.

\item[] Nonstrings may be produced from strings not only by multiplication but 
also by addition, as Figure (b) illustrates. We call pomset addition {\em 
concurrence}.

\item[] [Pratt goes on to define a number of other relations and apply them to a 
pomset-oriented model of computation.]

\end{itemize}
}

First an historical note. So far as I am aware, this is the first occurrence of 
the idea in computing theory: Smolensky [Smo90] apparently rediscovered it as 
the way to extend neural nets to symbols; it came to me, again independently, as 
the key to a distributed hierarchy concept in 1991.\footnote{In spite of 
superficial similarities, a phase web is quite different from a neural net. 
Also, [Smo90] has no hierarchical moment.} In fact, it is essentially the vector 
outer product originally discovered by Hamilton in 1825. Gibbs replaced it with 
the rather less fertile vector cross product late in that same century.

The above passage gives birth to the following comments:
\begin{itemize}
\item The multiplication $\times$ is {\em non-commutative}, in that the sequence 
(eg.) $TR$ - Transmit then Receive - is clearly not equivalent to $RT$ - Receive 
then Transmit. This is a principal characteristic of the outer product and gives 
rise to the appearance of $i=\surd -1$, which in turn is bound up with the 
rotation group. State change \underline{in general} can be expressed as a 
rotation in a vector space, and in this connection note that the state 
transformations in (a) can be viewed as successive 90$^o$ rotations [Note 2]. 
This notion that has been broadly exploited in both quantum and relativistic 
physics in this century because of its cogency. The phase web model does the 
same, and moreover in the context of {\em hierarchies} of vector spaces.

\item In fact {\em three} different sequences are provided for: two via 
successive left or right 90$^o$ rotations, and a third - which states/requires 
no ordering on the occurrence of $(1,T)$ and $(0,R)$ - as a direct 180$^o$ 
rotation.

\item The ``need for pomsets rather than posets'' brings up the issue of 
indistinguishability, a profound concept that has played a prominent role in the 
present author's work: concurrent events are namely indistinguishable in time.

\item ``Actions label events'' brings up the issues of how actually to both 
discover and label collections of events deserving the connotations of the 
concept of `action'. The discovery process is important in the present model, 
since this is required of a self-organizing system. In connection with both this 
and the preceding item, see Bastin \& Kilmister and Parker-Rhodes in the 
references.

\item ``We call pomset addition {\em concurrence}''. In the following we use $+$ 
to denote the concurrence of actions or states\footnote{In that our entities 
form groups [Man94], and a group element (state) can also be an operator 
(action).} - hereafter termed {\em co-occurrence} - which turns out to 
correspond directly to term-addition in a Clifford algebra.
\end{itemize}

Having introduced the basic theme of this paper, we leave Pratt at this point to 
pursue a different continuation, one which as noted in the introduction, eschews 
sets and strings of events in favor of other mechanisms. It is our feeling that 
orthocurrence failed to catch on because the full power of the concept, as well 
as the depth of its conceptual underpinnings, is difficult to see in the pomset 
context. We therefore examine more closely first co-occurrence, and thereafter 
the action-discovery process. We then present in rough outline the path to 
algebraic topology and its use in describing distributed computations 
hierarchically.

\section{Escaping from Turing's Box}\label{Co-occ}

An implicit claim of the Turing model is that a single sequence of 
computational events can capture all essential aspects of computation, that is, 
that {\em computation consists only of state \underline{transformations}}. To 
refute this claim, consider the following gedanken experiment:

{\em {\bf The coin demonstration} - Act I. A man stands in front of you with 
both hands behind his back, whilst you have one hand extended in front of you, 
palm up. You see the man move one hand from behind his back and place a coin on 
your palm. He then removes the coin with his hand and moves it back behind his 
back. After a brief pause, he again moves his hand from behind his back, places 
what appears to be an identical coin in your palm, and removes it again in the 
same way. He then asks you, ``How many coins do I have?''.}

It is important at the outset to understand that the coins are {\em formally} 
identical: indistinguishable in every respect. If you are unhappy with this, 
replace them with electrons or geometric points. Also, there are no `tricks' in 
the prose formulation. What is at issue is the fact of indistinguishability, and 
we are simply trying to pose a very simple situation where it is 
indistinguishability, and nothing else, that is in focus.

The indistinguishability of the coins now agreed, the most inclusive answer to 
the question is ``One or more than one'', an answer that exhausts the universe 
of possibilities given what you have seen, namely {\em at least} one coin. 
There being exactly two possibilities, the outcome can be encoded in one bit of 
information. Put slightly differently, when you learn the answer to the 
question, you will per force have received one bit of information.

{\em {\bf The coin demonstration} - Act II. The man now extends his hand and 
you see that there are two coins in it. [The coins are of course identical.]}

You now know that there are two coins, that is, {\em you have received one bit 
of information.} We have now arrived at the final act in our little drama.

{\em {\bf The coin demonstration} - Act III. The man now asks, ``Where did that 
bit of information come from??''}

Indeed, where {\em did} it come from?! Since the coins are indistinguishable, 
seeing them one at a time will never yield an answer to the question. Rather, 
{\em the bit originates in the simultaneous presence of the two coins}. We call 
such a confluence a {\em co-occurrence}. In that a co-occurrence, by 
demonstration a bona fide computational entity, is `situational' rather than 
`transformational', the above assumption that computation is purely 
transformational is shown to be false.

Addressing briefly the most common objections:

Q: Whatever you do, it can be simulated on a TM.\\
A: You can't `simulate' co-occurrence sequentially, cf. the Coin demo.

Q: But you can only check for co-occurrence sequentially - there's always a 
$\Delta t$.\\
A: This is a technological artifact: think instead of constructive/destructive 
interference - a phase difference between two wave states can be expressed 
in one bit.

Q: Simply define a TM that operates on the two states as a whole - the 
``problem'' disappears.\\
A: This amounts to an abstraction, which hierarchical shift changes the 
universe of discourse but doesn't resolve the limitation, since one can ask 
this new TM to `see' a co-occurrence at the new level. This objection thus 
dodges the issue.

Q: Co-occurrence is primitive in Petri nets, but these are equivalent to finite 
state automata.\\
A: The phase web in effect postulates {\em growing} Petri nets, both in nodes 
and connections. All bets are then off.

At this juncture, we hasten to mention that we are dealing here with {\em 
local} simultaneity, so there is no collision with relativity theory. Indeed, 
Feynman [Feyn65 p.63] argues from the basic principle of relativity of 
motion, and thence Einstein locality, that if {\em anything} is conserved, it 
must be conserved {\em locally}; see also [Man92, Phipps, Pope\&Osborne].

Penrose [Pen89] has argued that computational systems, not least parallel ditto, 
{\em in principle} cannot model quantum mechanics. However, we believe that his 
argument, together with most research involving (namely) parallelism, is subtly 
infected with the sequential mind-set, going back to Turing's analysis, and 
truly, earlier. A deep analogy with the difference between Newtonian and 
20$^{th}$ century physics is entirely defensible. The Coin demonstration is our 
reply to such arguments, which we do not then expect to hold.

Notice by the way how the matrix-based formulations of QM neatly get around 
the inherent sequentiality of $y=f(x)$-style (ie. algorithmic) thinking, namely 
by the literal co-occurrence of values in the vectors' and matrices' very 
layouts; and thereafter by how these values are composed {\em simultaneously} 
(conceptually speaking) by matrix operations. Instead of the matrix route, we 
have taken the conceptually compatible one of Clifford algebras, which are much 
more compact, elegant, and general, cf. [Hestenes].

Returning to our discussion of Turing's model, we see from the Coin 
demonstration that there is information, {\em computational information}, 
available in the universe {\em which {\bf in principle} cannot be obtained 
sequentially.} Replacing the coins by synchronization tokens, we can say that 
the information received from observing a co-occurrence is indicative of the 
fact that two states do not mutually exclude each other. 

Co-occurrence and mutual-exclusion are in fact conceptual {\em opposites}, in 
that (say) two events cannot simultaneously both co-occur and mutually exclude. 
The following section shows how this insight can be promoted to a concept of 
`action'.

\newpage
\section{Co-Exclusion}

{\em {\bf The block demonstration}. Imagine two `places', $p$ and $q$, each of 
which can contain a single `block'. Each of the places is equipped with a 
sensor, $s_{p}$ respectively $s_{q}$, which can indicate the presence or 
absence of a block.}

The sensors are the {\em only} source of information about the state of their 
respective places and are assumed {\em a priori} to be independent of each 
other, though they may well be correlated. The two states of a given sensor $s$ 
are mutually exclusive, so a place is always either `full', denoted 
(arbitrarily) by $s$, or `empty', denoted by $\tilde{s}$; clearly, 
$\tilde{\tilde{s}}=s$.\footnote{We are working in $\Bbb{Z}_{3}= 
\{0,1,-1=\tilde{1}\}$ rather than the traditional $\Bbb{Z}_{2}=\{0,1\}$. We use 
the visual convention that a sensor written without a tilde is taken to be bound 
to the value $1$, and vice versa; clearly, $\tilde{0}=0$.}

{\em Suppose there is a block on $p$ and none on $q$. This will allow us to 
observe the co-occurrence $s_{p} + \tilde{s}_{q}$. From this we learn that 
having a block on $p$ does not exclude not having a block on $q$. Suppose at 
some other instant (either before or after the preceding) we observe the 
opposite, namely $\tilde{s}_{p} + s_{q}$. We now learn that not having a 
block on $p$ does not exclude having a block on $q$. What can we conclude?}

First, it is important to realize that although the story is built around the 
co-occurrences $s_{p} + \tilde{s}_{q}$ and $\tilde{s}_{p} + s_{q}$, 
everything we say below applies equally to the `dual' pair of co-occurrences 
$s_{p} + s_{q}$ and $\tilde{s}_{p} + \tilde{s}_{q}$. [Note 1.] After all, the 
designation of one of a sensor's two values as `$\sim$' is entirely arbitrary. 
It is also important to realize that the places and blocks are story props: all 
we really have is two two-valued sensors reflecting otherwise unknown activities 
in the surrounding environment. Such sensors constitute the {\em boundary} 
between an entity and its environment in the phase web paradigm.

Returning to the question posed, we know that $s_{p}$ excludes $\tilde{s}_{p}$ 
and similarly $s_{q}$ excludes $\tilde{s}_{q}$. Furthermore, we have observed 
the co-occurrence of $s_{p}$ and $\tilde{s}_{q}$ and vice versa. Since the 
respective parts of one co-occurrence exclude their counterparts in the other 
co-occurrence (cf. first sentence), we can conclude that the co-occurrences 
{\em as wholes} exclude each other.

Take this now a step further. The transition $s_{p} \rightarrow \tilde{s}_{p}$ 
is indicative of some {\em action} in the environment, as is the reverse, 
$\tilde{s}_{p} \rightarrow s_{p}$. The same applies to $s_{q}$. Perceive the 
transitions $s_{p} \leftrightarrow \tilde{s}_{p}$ and $s_{q} \leftrightarrow 
\tilde{s}_{q}$ as two sequential computations, each of whose states consists of 
a single value-alternating bit. By the independence of sensors, 
these two computations are completely independent of each other. At the same 
time, the logic of the preceding paragraph allows us to infer the existence of 
a third computation, a {\em compound} action, with the state transition 
$s_{p} + \tilde{s}_{q} \leftrightarrow \tilde{s}_{p} + s_{q}$, denoted 
$s_{p}s_{q}$. In effect, by 
combining in this way two single-bit computations to yield one two-bit 
computation, we have lifted our conception of the actions performable by the 
environment to a new, higher, level of abstraction. This inference we call {\em 
co-exclusion}, and can be applied to co-occurrence pairs of any arity $>1$ 
where at least two corresponding components have changed.\footnote{The term 
`inference' is to be taken in its generic, not its formal logical, sense: 
co-exclusion is more nearly inductive in its thrust.}

Notice that the same reasoning applies to the action $s_{p}+s_{q} 
\leftrightarrow \tilde{s}_{p} + \tilde{s}_{q}$, also denoted $s_{p}s_{q}$ . The 
two actions are, not surprisingly, {\em dual} to each other, so co-exclusion on 
two sensors can generate two distinct actions. Like co-occurrence, an action 
defined by co-exclusion also possesses an emergent property, generally 
comparable to spin $\frac{1}{2}$ [Man94].

Co-exclusion provides a very general way for an entity to self-assemble: simply 
observe co-excluding co-occurrences, since these then will represent an 
abstraction of the environment. However, the mechanism for actually discovering 
co-exclusions is as yet unspecified. Speaking now very computationally, how {\em 
exactly} does one discover the existence of a co-exclusive relationship between 
two co-occurrences? 

Were computational resources not an issue, one could simply create a process to 
look for each possible co-exclusion - eg. first the one co-occurrence, and 
thereafter the other. When both have been observed, the existence of the 
co-exclusion is established, and a corresponding action can be instantiated. 
However, for $n$ sensors this requires $\cal{O}(2^n)$ processes! One could of 
course pre-specify which co-occurrences should be examined, but this eliminates 
the crucial element of discovery - one can hardly call a system {\em 
self}-organizing when one has more or less pre-specified how it is to put things 
together.

Rather than this, suppose we define a co-occurrence in terms of an ``event 
buffer'' with time-window-size $\Delta t$, where true simultaneity requires that 
$\Delta t = 0$, and larger values recognize the factual granularity with which 
the hardware can resolve events and/or the time-scale at which an entity 
interacts with its environment. Suppose further that event identifiers are put 
into the event buffer as they occur, ie. the new state engendered (and labelled 
by) the action associated with the event is inserted into the buffer. Finally, 
suppose that events in the buffer are successively discarded as their residence 
exceeds $\Delta t$ (or the same event-state changes again). Clearly, this 
arrangement guarantees that the state changes contained in the buffer all took 
place within $\Delta t$, and thus  occurred `simultaneously' (modulo $\Delta 
t$). The reader is at this point encouraged to ponder the fact that this 
mechanism in fact solves the problem of discovering co-exclusions, and at that 
in linear time and space, and without pre-specification! [Reader pause, for a 
lovely {\em aha!} experience.]

To see why this claim is true, consider the fact that a sensor's states are 
mutually exclusive, that is, if it is currently $s$ then before it changed it 
was in the state $\tilde{s}$. Furthermore, in $Bbb{Z}_3$ at least, the opposite 
is also true: $\tilde{\tilde{s}}=s$. Hence, since the buffer contains the 
co-occurrence (say) $s_{1}+s_{2}$, and {\em they both just changed}, then before 
they entered the buffer, $\tilde{s}_{1}+\tilde{s}_{2}$ obtained. But these two 
co-occurrences are exactly those necessary to define the co-exclusion 
$s_{1}+s_{2} \longleftrightarrow \tilde{s}_{1}+\tilde{s}_{2}!$ The computation 
time and space are fundamentally linear because they are proportional to the 
buffer size. If we specify that {\em all} events are to pass through our event 
buffer, then the only pre-specification is the arity of the co-exclusion. Even 
this pre-specification can be avoided if all possible co-exclusions (over the 
current buffer contents) are instantiated as each event is entered into the 
buffer. In all cases, a simple (commutative) hash-check of the co-exclusion's 
components can reveal duplicates [ManUS].

\section{Computation via Clifford Algebras}

This section presents, very informally, the mathematical foundation of the phase 
web paradigm. The point of departure is to view sensor states as vectors instead 
of scalars, as is conventionally done. 

Let sensor state $s=1$ indicate that sensor $s$ is  currently being 
stimulated, ie. a synchronization token for that state is present, and 
$s=\tilde{1}$ that $s$ is currently {\em not} being stimulated, and 
hence a token for state $\tilde{s}$ is present. Thus the two states of $s$ are 
represented by their respective synchronization tokens, whose respective 
presences by definition exclude each other.

That the sensors {\em qua} vectors are orthogonal derives from the fact that, 
in principle, a given sensor says nothing about the state of any other sensor. A 
state of a multi-sensor system is then naturally expressed as the sum of the 
individual sensor vectors, and the state $(s_{a},\tilde{s}_{b}) = (1,\tilde{1})$ 
is written as the vector sum $s_{a}+\tilde{s}_{b}$. [Note 2.] Since such states 
represent co-occurrences, it follows that co-occurrences are vector sums, 
usually denoting partial (local) states. Note how the commutativity of `$+$' 
reflects the lack of ordering of the components of a co-occurrence; and as well 
that the co-occurrence $1+\tilde{1}=0$ indicates that the interpretation of 
`zero' is that the components of the sum {\em exclude} each other. Because 
$\Bbb{Z}_{2}$ does not distinguish state-value and exclusion, we take our 
algebra to be over $\Bbb{Z}_{3} = \{0,1,2\}= \{0,1,\tilde{1}\}$.

The next step is to represent {\em actions}. [Man94] presents a detailed 
analysis of the group properties of co-occurrences and actions, concluding that 
the appropriate algebraic formalism is a (discrete) Clifford algebra, and that 
the state transformation effected by an action is naturally expressed using this 
algebra's vector product. A prime characteristic of this product is that it is 
anti-commutative, that is, for $(s_{1})^2 = (s_{2})^2 = 1$, $s_{1}s_{2} = 
-s_{2}s_{1}$.\footnote{The Clifford product $ab$ can be defined as $ab=a\cdot b 
+ a\wedge b$, ie. the sum of the inner ($\cdot$) and outer ($\wedge$) products, 
where $a\wedge b = -b\wedge a$ is the oriented area spanned by vectors $a,b$. 
The basis vectors $s_{i}$ of a Clifford algebra may have $(s_{i})^2=\pm 1$, and 
while here we choose $+1$, reasons are appearing for choosing $-1$. As long as 
they all have the same square, it doesn't matter for what is said here. Note 
that $(s_{1}s_{2})^{2}=-1$, so $s_{1}s_{2} \cong \surd -1$, cf. \S 1.} The 
magnitude of any such product is the area of the parallelogram its two 
components span, and the {\em orientation} of the product is perpendicular to 
the plane of the parallelogram and determined by the ``right hand rule''. 
Applying the Clifford product to a state, one finds - using the square-rule and 
the anti-commutativity of the product given above - that 
\begin{equation}
(s_{1}+s_{2})s_{1}s_{2} = s_{1}s_{1}s_{2} + s_{2}s_{1}s_{2} = s_{2} + 
\tilde{s}_{1}s_{2}s_{2} = \tilde{s}_{1} + s_{2}
\label{(1)}
\end{equation} 
that is, that the result of the {\em action} $s_{1}s_{2}$ is to rotate the 
original state by $90^{o}$, for which reason things like $s_{1}s_{2}$ are called 
{\em spinors}. Thus {\em state change} in the phase web is modelled by rotation 
(and reflection) of the state space, and the effect of an `entire' action can be 
expressed by the inner automorphism $s_{1}s_{2}(s_{1} + s_{2})s_{2}s_{1} = 
\tilde{s}_{1}+\tilde{s}_{2}$, which corresponds to a rotation through $180^o$.
[Note 3.] 
Using this automorphism, one can derive two families of action-composition rules 
(here, for arity 2):
$$s_{1}s_{3}(s_{1}+s_{2}+s_{3})s_{3}s_{1}= 
[s_{2}s_{3}]s_{1}s_{2}(s_{1}+s_{2}+s_{3})s_{2}s_{1}[s_{3}s_{2}] = 
\tilde{s}_{1}+s_{2}+\tilde{s}_{3}$$
which is a traditional functional composition, and the parallel composition 
$$s_{1}s_{3}(s_{1}+s_{3})s_{3}s_{1} = s_{1}s_{2}(s_{1} + s_{2})s_{2}s_{1} + 
s_{2}s_{3}(\tilde{s}_{2} +  s_{3})s_{3}s_{2}$$

One of the felicities of Clifford algebras is that one needn't designate one of 
the axes as `imaginary' and the other as `real'. Rather, the $i$-business is 
implicit  and the  algebra's anti-commutative product neatly bookkeeps the 
desired orthogonality and inversion relationships, no matter how many dimensions 
[ie. sensors (roughly)] are present. The action-as-product and its implicit $i$ 
implements the transformations necessary to preserve consistent 
observer/observed relationships; this will be clearer when we discuss the 
hierarchical model later.

The above 2-spinors are just one example of the vector products available in a 
Clifford algebra - any product of the basis vectors $s_i$ is well-defined, and 
just as $s_{1}s_{2}$ defines an area, $s_{1}s_{2}s_{3}$ defines a volume, etc.  
Being by nature mutually orthogonal, the terms of a Clifford algebra 
\begin{equation}
s_{i} + s_{i}s_{j} +  s_{i}s_{j}s_{k} + \dots + 
s_{i}s_{j}\dots s_{n}
\label{(2)}
\end{equation} 
themselves also define a vector space, which is the space in which we will be 
working (actually, hierarchies of such spaces). [The term (eg.) $s_{i}s_{j}$ 
above, for $n=3$, denotes $s_{1}s_{2}+ 
s_{2}s_{3}+s_{1}s_{3}$, that is, all possible non-redundant combinations.] It is 
perhaps worth stressing that this vector space is the space of the {\em 
distinctions} expressed by sensors, and as such has no direct relationship with 
ordinary 3+1 dimensional space.

A Clifford product like $s_{1}s_{2}$ reflects both (1) the emergent aspect of a 
phase web action (via its perpendicularity to its components) and (2) its 
ability to act as a meta-sensor (since its orientation is $\pm 1$). Regarding 
(1), the emergence is rooted in the information gleaned from the co-occurrences 
underlying the co-exclusion inference that yields $s_{1}s_{2}$, cf. the Coin 
demonstration. Regarding (2), the co-exclusion inference is an {\em abstraction} 
that produces a single action with two bits of state from two lower level 
actions each possessing a single bit of state. Since this abstraction has the 
same external behavior as its constituent sensors, namely $\pm 1$, we can 
legitimately view it too as a sensor, a {\em meta-}sensor. By co-excluding 
meta-sensors, we can build a new set of abstractions - meta-meta-sensors - etc., 
and thus construct a hierarchy of interwoven co-occurrences and exclusions that 
directly reflects the {\em observed} structure of the surrounding environment. 
This hierarchy is the topic of the following section.

\section{Extension to Hierarchical Spaces}

This section introduces the hierarchical aspect of the phase web model of 
computation. It is fitting that we begin with the question ``why hierarchy?''.

The concept of hierarchy has been used in computing for a long time - the use of 
a hierarchical structure, with its accompanying logarithmic reduction in the 
number of entities to be juggled, is a major key to aesthetic, cogent, and 
maintainable programs. The current situation in computing - where computational 
models of other than computer systems are becoming widespread - offers several 
new reasons to look more closely at hierarchy:

\begin{itemize}
\item Contemporary physical theory is unable to tell a convincing story of how 
one gets from its microscopic conceptual foundation - quantum mechanics - to the 
phenomena of the macroscopic world of elbow joints and eyes, evolution and 
ecosystems, and the thermodynamic arrow of time (irreversibility), to name a 
few.

\item The continued miniaturization of hardware components is leading to 
increasing contact with the quantum mechanical world, yet our understanding of 
computation is strongly Newtonian (total orders, billiard-ball causality, etc.). 
This hierarchically `downward' extension of our understanding is an inverse 
formulation of the preceding problem.

\item Living systems can in fact {\em not} be described with physical theories 
in their current form [Ros85, Ros91, Mat89]. In addition, the obviously 
hierarchical structure of multi-celled organisms, not to mention DNA-driven 
assembly processes, are quite untouched by contemporary research into 
multi-process computational systems. The computational description of 
self-organization mentioned earlier is another example.
\end{itemize}

The generalization of function composition hierarchy to concurrency and 
distribution has proven elusive. This generalization is critical, however, if 
computational concepts are to be applied to natural systems. As a case in point, 
computing has long rested on the assumption that the behavior of a computation 
was independent of its physical substrate, but the complexity results from 
quantum computation show that something important is missing.
Looking at the matter hierarchically, the issue is whether/how can one descend 
through levels from a macroscopic description to a microscopic one and arrive at 
quantum mechanics. What Penrose in effect argued is that if you use a 
function-composition hierarchy - which has a hand and glove relationship to 
Turing's model - you'll never get there. The ever-shrinking size of circuit 
components to a scale where quantum effects are unavoidable thus represents a 
genuine crisis. It is therefore comforting that the phase web's underlying 
mathematical structure is entirely compatible with both relativity and quantum 
theory. For example,
\begin{itemize}
\item Phase web computations inherently superpose states as sub-goals pursue 
alternative paths breadth-first to achieve the given goal - a classic quantum 
behavior.

\item 2- and 3-spinors exhibit spin $\frac{1}{2}$ - also a characteristic 
quantum phenomenon; and co-occurrences spin 1. The isomorphism between the two 
sides of Figure \ref{BowdenBase} is strikingly analogous with the thrust of 
contemporary string theory - that there is a 1-1 relationship between fermions 
and bosons.

\item The `objects' created by co-exclusion satisfy a mutex-based resource 
invariant, which we hypothesize are the computational analog of quantum 
conservation laws [Man92].
\end{itemize}
In addition, [Man97] shows how living systems can be modelled by the methods we 
describe here. It is the contention of this article that a hierarchical view of 
distributed computational systems is the key, not only to a deeper understanding 
of the nature and depth of the very notion of computation itself, but also to 
the solution of the above problems. We should therefore begin to {\em insist} 
that contemporary theories of computation demonstrate their applicability to 
such larger scientific issues, in addition to traditional, strictly 
computational, concerns.

\subsection{The Twisted Isomorphism}

One might expect that the co-exclusion of two meta-sensors, say 
$s_{i}s_{j}$ and $s_{p}s_{q}$, would be modelled by simply multiplying them, to 
get the 4-action $s_{i}s_{j}s_{p}s_{q}$. This turns out however to be 
inadequate, since although by the same logic the co-exclusion of (say) $s_{i}$ 
and $s_{i}s_{j}$ in Topsy expresses explicitly a useful relationship (eg. 
part-whole), the algebra's rules reduce it from $s_{i}s_{i}s_{j}$ to $s_j$, 
which is simply redundant.

Instead, we take as a clue the fact that goal-based {\em change} in Topsy 
occurs via trickling down through the layers of hierarchy, and draw an analogy 
with differentiation. In the present decidedly geometric and discrete context, 
differentiation corresponds to the {\em boundary operator} $\partial$. Define 
$\partial s = 1$ and let
$$\partial(s_{1}s_{2}\dots s_{m}) = 
s_{2}s_{3}\dots s_{m} - s_{1}s_{3}\dots s_{m} + s_{1}s_{2}s_{4}\dots s_{m} - 
\dots (-1)^{m+1}s_{1}s_{2}\dots s_{m-1}$$ 
that is, drop one component at a time, in order, and alternate the sign. [Note 
4.] Using the algebra's rules as before, one can show that 
$\partial(s_{1}s_{2}\dots s_{m}) = (s_{1}+s_{2}+\dots+s_{m})s_{1}s_{2}\dots 
s_{m}$ which is exactly the form of equation (1) for what an action does!

Take now equation (2) expressing the vector space of distinctions, segregate 
terms with the same arity, and arrange them as a decreasing series:
\begin{equation}
 s_{i} \stackrel{\partial}{\longleftarrow} s_{i}s_{j} 
\stackrel{\partial}{\longleftarrow} s_{i}s_{j}s_{k} 
\stackrel{\partial}{\longleftarrow} \dots \stackrel{\partial}{\longleftarrow}  
s_{i}s_{j}\dots s_{n-1} \stackrel{\partial}{\longleftarrow} s_{i}s_{j}\dots 
s_{n}
\label{(3)}
\end{equation}
Here as before, $s_{i}s_{j}$ is to be understood as expressing all the 
possible 2-ary forms (etc.), and hence the co-occurrence of pieces of 
similar structure. Each of the individuals is a {\em simplicial complex}, and 
the whole sequence is called a {\em chain complex}, expressing a sequence of 
structures of graded geometrical complexity in which the transition from a 
higher to a lower grade is defined by $\partial$. Furthermore, the entities at 
adjacent levels are related via their group properties - their {\em homology}, 
which we here assume is trivial.

It turns out that there is a second structure - a {\em cohomology} - that is 
isomorphic to the homology, but with the difference that arity  {\em increases} 
via the $\delta$ (or {\em co-boundary}) operator,\footnote{More precisely, 
$(\sigma_{p},\delta d^{p-1})=(\sigma_{p}\partial,d^{p-1})$, where $\sigma_{p}$ 
is a simplicial complex with arity $p$, and $d^{p}$ the corresponding 
co-complex. The isomorphisms $\mu,\mu^{-1}$ are matrices containing the terms' 
$\Bbb{Z}_{3}$ coefficients.} precisely opposite to $\partial$, cf. eqn. (3):
\begin{equation}
s_{i} 
\stackrel{\delta}{\longrightarrow} s_{i}s_{j} 
\stackrel{\delta}{\longrightarrow} s_{i}s_{j}s_{k} 
\stackrel{\delta}{\longrightarrow} \dots \stackrel{\delta}{\longrightarrow}  
s_{i}s_{j}\dots s_{n-1} \stackrel{\delta}{\longrightarrow} s_{i}s_{j}\dots 
s_{n}
\label{(4)}
\end{equation}
Building such increasing complexity is exactly what co-exclusion does. [We note 
that a Clifford algebra satisfies the formal requirements for the existence of 
the associated homology and cohomology.]

It is easily proven that $\partial\partial=0$, and by isomorphism, so also 
$\delta\delta=0$. For example, $\partial\partial(s_{1}s_{2}) = 
\partial(\tilde{s}_{1}+s_{2}) = \tilde{1}+1 = 0$, and similarly, $\partial 
\partial (s_{2}s_{1}) = \partial(s_{1}+\tilde{s}_{2}) = 1+\tilde{1} = 0$. 
Combining these now as the exclusion $\partial\partial(s_{1}s_{2}+s_{2}s_{1})$, 
we get $(1+ \tilde{1}) + (\tilde{1}+1) = (1+1)+(\tilde{1}+\tilde{1}) = 0$,
which are the two forms of the input to the determination of a co-exclusion 
relationship. Recalling the event-buffer mechanism for discovering 
co-exclusions, we see, especially if $\Delta t=0$, that this mechanism is a 
realization of the isomorphic $\delta\delta=0$ !

Viewing $\delta$'s abstraction operation informationally, we see that two bits 
($s_{1},s_{2}$) are being encoded in a single bit (the orientation of 
$s_{1}s_{2}$), that is, information is being `abstracted away'. The missing bit 
indicates the {\em phase} of the action, eg. whether the state 
rotation/transformation is $s_{1}+s_{2} \leftrightarrow 
\tilde{s}_{1}+\tilde{s}_{2}$ or $s_{1}+\tilde{s}_{2} \leftrightarrow 
\tilde{s}_{1}+s_{2}$. What will actually occur is however well-defined by the 
other connections $s_{1},s_{2}$ partake in, ie. the boundary conditions of the 
action. Note however that `well-defined' does not necessarily imply 
`deterministic', cf. the earlier comment on 90$^o$ vs. 180$^o$ rotations. 
Isomorphically, the corresponding $\partial$ operation destroys the emergent 
information in the current state and replaces it by non-deterministic choice.  

\begin{figure}[htbp]
\begin{center}
\leavevmode
\epsfbox{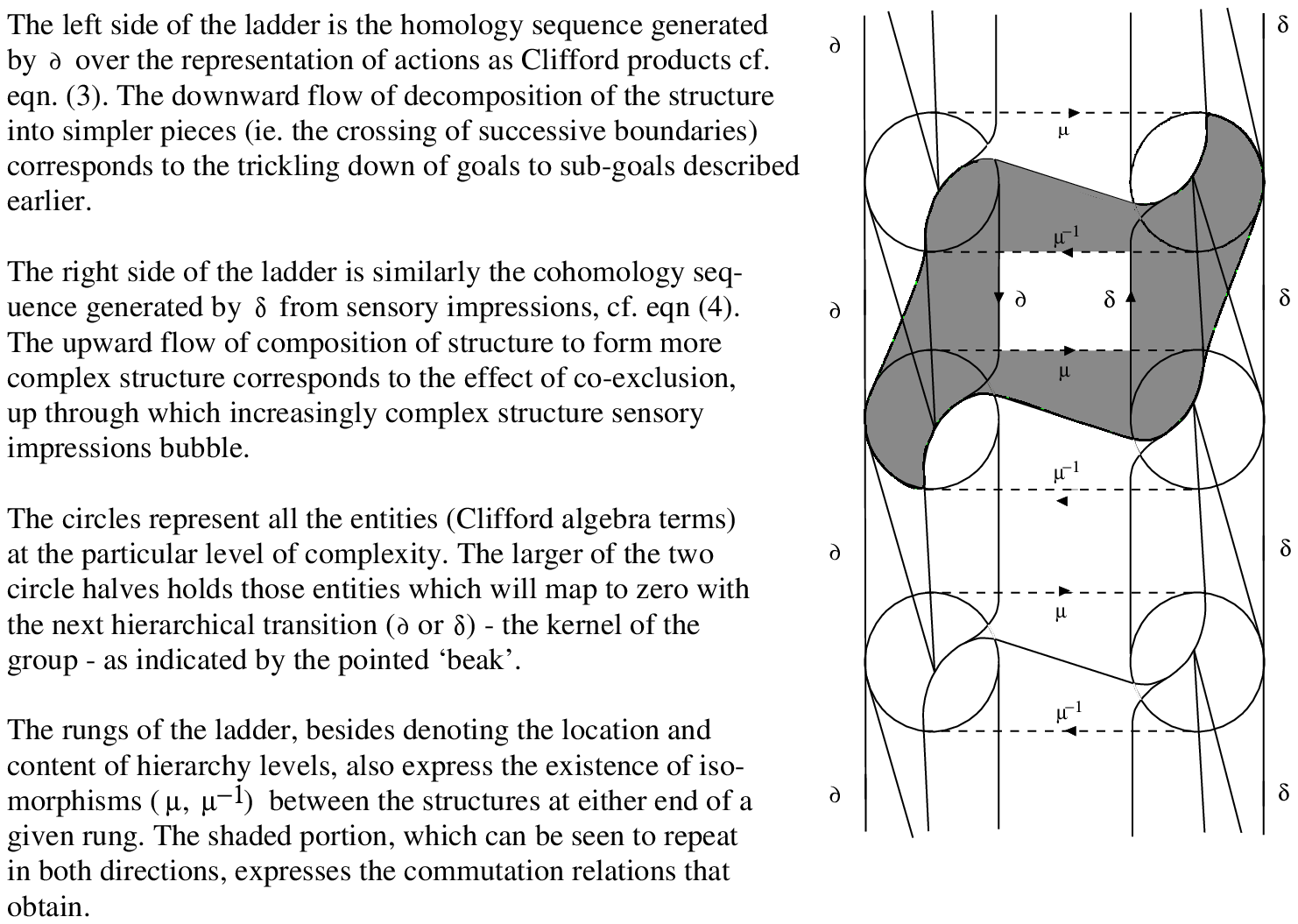}
\caption{Ladder diagram, illustrating homology-cohomology relationships.}
\label{BowdenBase}
\end{center}
\end{figure}

Refer now to Figure \ref{BowdenBase} [Bow82], which we call a {\em ladder 
diagram}.\footnote{Strictly speaking, $\partial, \delta$, and $\mu/\mu^{-1}$ 
should be indexed by level: $\partial_{\ell}, \delta_{\ell}, 
\mu_{\ell}/\mu^{-1}_{\ell}$.} 

The shaded shape points out a unique property of the homology-cohomology 
ladder, one that even many topologists seem unaware of, namely that the 
isomorphisms $\mu, \mu^{-1}$ are {\em twisted}, that is, the kernel of the 
group at one end of a rung is mapped by $\mu$ (respectively, $\mu^{-1}$) into 
the non-kernel elements of the group at the other end. This property was 
discovered by [Roth] in his proof of the correctness of Gabriel Kron's then 
controversial methods for analyzing electrical circuits [Bow82], and turns 
out to have profound implications: the entirety of Maxwell's 
equations and their interrelationships can be expressed by a ladder with two 
rungs plus four terminating end-nodes [Bowden], and [Tonti] has - 
independently - shown similar relationships for electromagnetism and 
relativistic gravitational theory. Roth's twisted isomorphism (his term) thus 
reveals the deep structure of the concept of boundary, and shows that the 
complete story requires both homology and cohomology.

\subsection{Generalizing the Twisted Isomorphism Hierarchy}

Each level of a ladder hierarchy, as presented so far, is built entirely from 
entities (ie. sensors) from the level immediately underneath, leading to what we 
call a `pancake' hierarchy. But this is an unnecessary limitation, from which we 
now generalize.

Let $S_{i}$ be the set of sensors at $\delta$-level $i$. Similarly, let $G_{i}$ 
be the set of (sensors expressing the presence of) goals at $\partial$-level 
$i$. A pancake {\em meta} hierarchy of 2-actions can now be characterized by 
$S_{i} = S_{i-1} \times S_{i-1}$, where $\times$ is the cartesian product 
mediated by $\delta$. Other, more general, hierarchical forms are now easily 
seen:

\begin{itemize}
\item $S_{i} = S_{j} \times S_{k}$, $j,k < i$, yielding non-pancake meta 
hierarchies; and of course the product may be over $>$2 levels. Aside from this, 
however, the semantics is roughly as before;

\item $G \times G$, yielding a purely goal-based {\em icarian} hierarchy, 
roughly similar to a function-composition hierarchy;

\item $S \times G$, yielding a combined abstraction over the underlying ladder 
level(s) that we call a {\em morphic} hierarchy.
\end{itemize}

\begin{figure}[htbp]
\begin{center}
\leavevmode
\epsfbox{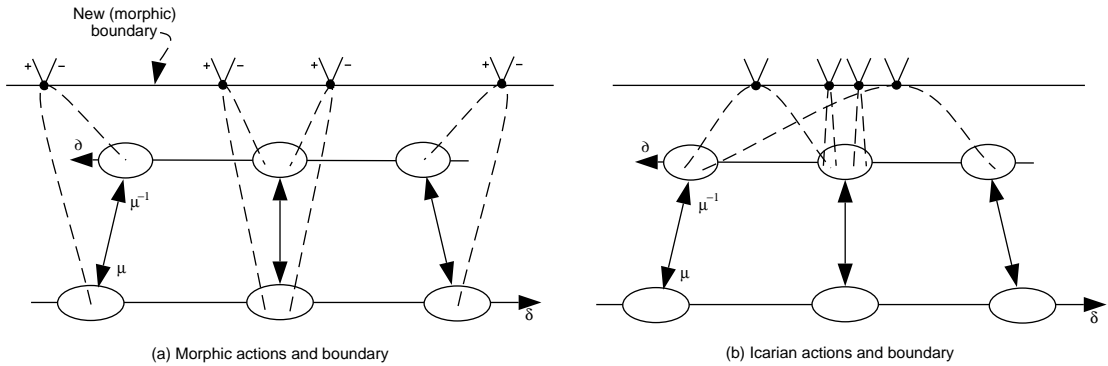}
\caption{Morphic vs. icarian hierarchies.}
\label{MIHier1}
\end{center}
\end{figure}

Figure \ref{MIHier1} illustrates the latter two, and we note that the morphic 
level in (a) may in principle `cross' levels more radically, eg. as (b) does.
We call these generalized forms {\em ortho-hierarchies}.

Icarian actions provide a means for a computation to express, self-reflectively, 
the {\em way} it carries out its goals. Morphic actions provide a means for a 
computation to express, self-reflectively, the relationship between $S$ and $G$ 
that is otherwise buried in $\mu,\mu^{-1}$. All three generalizations are 
supported by Topsy.

\newpage
\subsection{Causality}

In our view, the concept of causality is intertwined with those of hierarchy, 
atomicity, and sequence: hierarchy {\em is} atomicity, atomicity (one level 
down) {\em is} sequence, sequence {\em is} exclusion.

Ordinary, everyday causality says that if the pre-condition $p$ of action $q$ is 
satisfied, and $q$ occurs, then $p$ caused $q$. The troubles begin when eg. (1) 
we may not assume that $q$ is atomic, since other processes can now in principle 
access and change $q$'s intermediate states; and (2) there are superposed 
states.

Since co-exclusion keys on the mutual exclusion of two compound states, {\em 
each of whose components also exclude}, the entity it produces represents a 
mutual exclusion of a particular kind, namely, an exclusion whose violation 
produces literal meaninglessness. Pauli's exclusion principle [that no two 
electrons may be in the same state at the same time] is an example, since 
without it the way electrons form shells cannot be understood, whereas with it 
everything makes sense. This is quite unlike the more-or-less arbitrary mutexes 
we introduce by hand into programs to ensure a desired behavior. The mutual 
exclusions captured by co-exclusion express an inherent definition of an inner, 
mutual, logical consistency that is not the slightest bit arbitrary.

Hence, presuming that $q$ above is atomic in this way, it itself simply cannot, 
simultaneously, perform any other transition than the one given to it (namely $p 
\rightarrow \tilde{p}$). The only other possibility is that the transition does 
not occur. {\em But}, this is only when viewed at the ladder-level at which $q$ 
is defined. If then we wish to discuss {\em simultaneously} both $q$ {\em and} 
its components at the next level down, we must do so explicitly, which means 
that we must treat the entire structure
$\partial_{i} + \mu_{i-1} + \delta_{i-1} + \mu^{-1}_{i}$.
The simple causality story is no longer adequate because there is an unavoidable 
{\em definitional} interplay between levels. But, we claim, the 
ladder-hierarchical tools we present here, eg. morphic abstraction for the above 
causal analysis, provide a framework wherein the issues can be formulated with 
precision: we can simultaneously and rigorously `observe' and treat $q$, its 
components, the interplay with other actions, and this entire whole. In 
contrast, it is unobvious how to do this using function composition hierarchy.

The overall moral is that one must always keep in mind which level{\em s} of 
hierarchy one has engaged conceptually. Only when a single level is in play can 
one hope to simplify to ordinary Newtonian causality and use (eg.) the action 
composition rules given earlier.

\section{Summary and Conclusion}

We have shown how Pratt's original insight, taken in a different direction, 
leads to a radically new way to describe distributed computations. The current 
function-oriented focus on strings of events is replaced by a new kind of 
hierarchical abstraction that is squarely based on the phenomena peculiar to 
concurrent systems. The result is a model of concurrent computation where the 
duality of state and action, the interplay of atomicity and hierarchy, and 
causality and sequence can be treated in an integrated and mathematically 
powerful framework. This same framework can also describe self-organizing and 
self-reflecting computations. 

Equally important is the fact that this strongly hierarchical approach can tell 
us how to reduce macroscopic algorithms to quantum-mechanical circuits, and how 
to design and produce micro-assembled quantum devices. It also offers a way to 
understand such extremely complex things as DNA-directed assembly and living 
systems, and in general, to connect computing to contemporary physical 
mathematics in a way that mathematical logic never can.

Very little of this is possible with today's models of computation, mired as 
they are in concepts that are fundamentally sequential. We think it's time to 
change horses!

\vspace{3.0mm}

{\bf Acknowledgements}. Thanks to IEEE for permission to reprint the Coin and 
Block demonstrations; to Luca Aceto and Uwe Nestmann; and especially to my 
students.

\vspace{0.5cm}
{\bf References}

Bastin, T. and Kilmister, C.W. {\em Combinatorial Physics}. World Scientific, 
1995. ISBN  981-02-2212-2.

Bastin, T. and Kilmister, C.W. ``The Combinatorial Hierarchy and Discrete 
Physics''. Int'l J. of General Systems, special issue on physical theories from 
information (in press).

Bowden, K. ``Physical Computation and Parallelism (Constructive Post-Modern 
Physics)''. Int'l J. of General Systems, special issue on physical theories 
from information (in press).

Bowden, K. {\em Homological Structure of Optimal Systems}. PhD Thesis, 
Department of Control Engineering, Sheffield University UK. 1982.

Feynman, R. {\em The Character of Physical Law}. British Broadcasting Corp. 
1965.

Gelernter, D. ``Generative Communication in Linda''. ACM Transactions on 
Programming Languages, 1985.

Herlihy, M. and Rajsbaum, S. ``Algebraic Topology and Distributed Computing - A 
Primer''. Lecture Notes in Computer Science, v. 1000, pp. 203-217. Springer 
Verlag, 1995.

Herlihy, M. and Shavit, N. ``A simple constructive computability theorem for 
wait-free computation''. In {\em  Proceedings 26th Annual ACM Symposium on 
Theory of Computing}, pp. 243-252, May, 1994.

Hestenes, D. and Sobczyk, G. {\em From Clifford Algebra to Geometric Calculus.} 
Reidel, 1989.

Hestenes, D. {\em New Foundations for Classical Mechanics}. Reidel, 1986. The 
first 40 pages contain a very nice, historical introduction to the vector 
concept and Clifford algebras.

Manthey, M. ``Synchronization: The Mechanism of Conservation Laws''. Physics 
Essays (5)2, 1992. 

Manthey, M. ``Toward an Information Mechanics''. Proceedings of the 3rd IEEE 
Workshop on Physics and Computation; D. Matzke, Ed. Dallas, November 1994. ISBN 
0-8186-6715X.

Manthey, M. Distributed Computation, the Twisted Isomorphism, and Auto-Poiesis. 
Proceedings of the 1$^{st}$ Int'l Conference on Computing Anticipatory Systems 
(CASYS97). Liege, Belgium; August 1997, D. Dubois, Ed.

Manthey, M. US Patents 4,829,450, 5,367,449, and others pending. The intent 
is to license freely (on request) to individuals  and research or educational 
institutions for non-profit use. See [www] for licensing information.

Matsuno, K. {\em Protobiology: Physical Basis of Biology.} CRC Press, Inc. 1989.
ISBN 0-8493-6403-5.

Parker-Rhodes, F. {\em Theory of Indistinguishables - A Search for 
Explanatory Principles Below the Level of Physics}. Reidel. 1981.

Phipps, T. ``Proper Time Synchronization''. Foundations of Physics, vol 21(9)  
Sept. 1991.

Penrose, R. {\em The Emperor's New Mind}. Oxford University Press, 1989. ISBN 
0-19-851973-7.

Pope, N.V. and Osborne, A.D. ``Instantaneous Relativistic 
Action-at-a-Distance''. Physics Essays, 5(3) 1992, pp. 409-420.

Pratt, V. ``Modelling Concurrency with Partial Orders.'' Stanford University 
Computer Science Department report STAN-CS-86-1113. Also: Int'l J. of Parallel 
Programming, v15,1; and Pratt's www home-page.

Pratt, V. ``Chu Spaces and their Interpretation as Concurrent Cbjects''. Lecture 
Notes in Computer Science, v. 1000, pp. 392-405. Springer Verlag, 1995.

Robert Rosen, {\em Anticipatory Systems}. Pergamon Press, 1985. ISBN 
0-08-031158-x.

Rosen, R. {\em Life Itself - A Comprehensive Inquiry into the Nature, Origin, 
and Fabrication of Life.} Columbia University Press, 1991. ISBN 
0-231-07564-2.

Roth, J.P. ``An Application of Algebraic Topology to Numerical Analysis: On the 
Existence of a Solution to the Network Problem''. Proc.  US National Academy of 
Science, v.45, 1955.

Smolensky, P. ``Tensor Product Variable Binding and the Representation of 
Symbolic Structures in Connectionist Systems''. Artificial Intelligence 46 
(1990) 159-216. Elsevier Science Publishers B.V. (North-Holland).

Tonti, E. ``On the formal structure of the relativistic gravitational theory''. 
Accademia Nazionale Dei Lincei, Rendiconti della classe di Scienze fisiche, 
matematiche e naturali. Serie VIII, vol. LVI, fasc. 2 - Feb. 1974. (In 
english.)

www. Various phase web and Topsy publications, including (soon) code 
distribution, are available via {\em www.cs.auc.dk/topsy}.

\newpage
{\small


{\bf Notes.}

0. In contrast to Linda, TLinda is not intended to be embedded in another, 
algorithmic, language but rather to be used as a pure coordination language. As 
such TLinda eg. contains only integer inc/dec, and no comparisons; extends 
Linda's traditional \Ss{out}, \Ss{in}, \Ss{rd} and \Ss{eval} with \Ss{co}, 
\Ss{notco} and like operations; and in the interests of speed (which is 
respectable) requires pre-declaration of tuples. 

1. It sometimes troubles people that the elements of the co-occurrence (say) 
$s_{p}+ \tilde{s}_{q}$ don't seem at all indistinguishable - on the 
contrary, $s_{p}$ is clearly distinct from $\tilde{s}_{q}$! The confusion is 
understandable, and derives from confounding the {\em value} of a sensor with 
the synchronization {\em token} that represents the fact that the value (= 
process state) obtains for the moment. The difference is clearer in the 
implementation, where the tokens for the respective states of the sensor 
processes $s_p$ and $\tilde{s}_q$ are represented by the tuples \Ss{[p,1]} and 
\Ss{[q,\~1]}, which tuples can be thought of as making precise exactly {\em 
which} state's token is being referred to. The processes accessing such tuples 
in fact know {\em a priori} the exact form of the tuple (ie. state) they are 
interested in, so no information is conveyed by accessing such tuples (which is 
as it should be, since synchronization must not convey information between 
processes). Summa summarum, the sensor values are not what are distinguished, 
but rather the tokens representing the associated sensor-process states, and 
these tokens are indistinguishable {\em in time}. 

2. Figure (a) shows a single sensor's states expressed as a vector, and (b) the 
way two such vectors can indicate a state, eg. the pre-condition of an action.

\begin{figure}[htbp]
\begin{center}
\leavevmode
\epsfbox{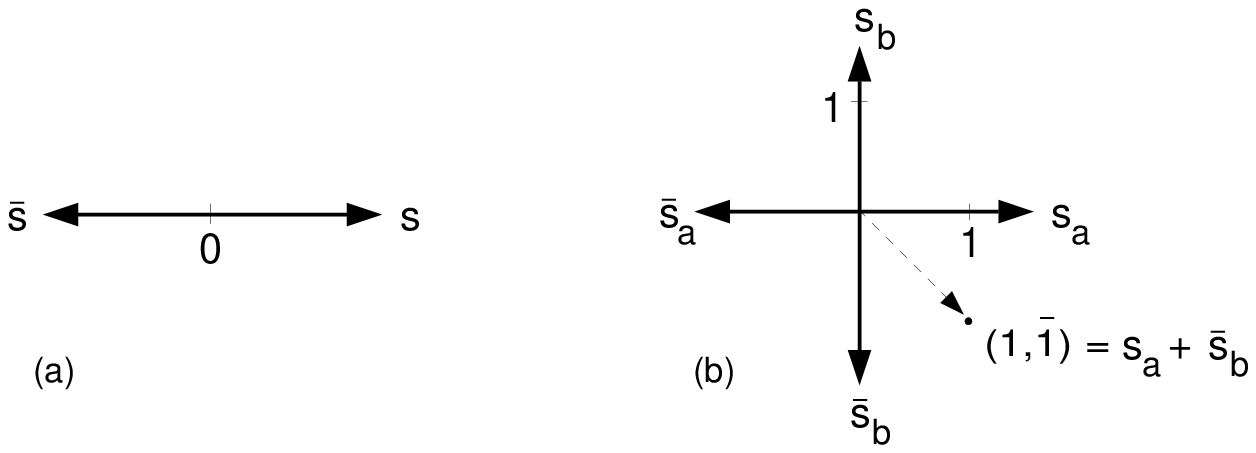}
\end{center}
\end{figure}
The mappings $R \mapsto s_{a}, T = \tilde{R} \mapsto \tilde{s}_{a}, 0 \mapsto 
s_{b}, 1 \mapsto \tilde{s}_{b}$ capture the connection to Pratt's example.

3. Taking the classic vending machine example, let $b$ be the button that 
selects coffee ($\mapsto b$) or tea ($\mapsto \tilde{b}$), $m$ the indicator of 
money present ($\mapsto m$) or not ($\mapsto \tilde{m}$), and $s$ the indicator 
that the beverage is served ($\mapsto s$) or not ($\mapsto \tilde{s}$); $s$ is 
reset to $\tilde{s}$ by removing the beverage from the machine. The action we 
want is $m+\tilde{s} \leftrightarrow \tilde{m}+s$, ie. $ms$. The automorphism 
then states that $ms(m+\tilde{s})sm = \tilde{m}+s$, ie. money together with no 
serving is transformed to no money and a serving. Invoking the lemma 
$x+ms(m+\tilde{s})sm = ms(x+m+\tilde{s})sm = x + \tilde{m}+s$, that is, $x$ 
($x=b=$coffee or $x=\tilde{b}=$tea) is present and unchanged throughout, then 
describes the vending machine's basic behavior.

4. The boundary operator $\partial$ has a straightforward geometric 
interpretation. A triangle $ABC$ specified in terms of its vertices $A,B,C$, has 
edges $AB, BC, CA$. Then $\partial(ABC) = BC - AC + AB$. Since specifying the 
triangle's edges in terms of its vertices means that edge $AC$ is oriented 
oppositely to edge $CA$, we can rewrite the above as $AB + BC + CA$, which is 
indeed the boundary of the triangle (versus its interior).

}
\end{document}